\documentclass[aps,pre,groupedaddress]{revtex4-2}
\usepackage{natbib}
\usepackage{amsmath}
\usepackage{epsfig}
\usepackage{color}
\usepackage{tikz}

\usepackage{graphicx}
\usepackage{amsfonts}% Include figure files
\usepackage{dcolumn}% Align table columns on decimal point
\usepackage{bm}% bold math
\usepackage{color}
\usepackage{hyperref}

\begin{document}
\bibliographystyle{unsrt}

\title{Stochastic Scovil--Schulz-DuBois machine and its three types of cycles}  
\author{Fei Liu} 
\email[Contact author: ]{feiliu@buaa.edu.cn}
\affiliation{School of Physics, Beihang University, Beijing 100083, China}

\author{Jiayin Gu}
\affiliation{School of Physics and Technology, Nanjing Normal University, Nanjing 210023, China}

\date{\today}

\begin{abstract}
Three types of cycles are identified in the quantum jump trajectories of the Scovil--Schulz-DuBois (SSDB) machine: an R-cycle as refrigeration, an H-cycle as a heat engine, and an N-cycle in which the machine is neutral. The statistics of these cycles are investigated via a semi-Markov process method. We find that in the large time limit, whether the machine operates as a heat engine or refrigerator depends on the ratio between the numbers of R-cycles and H-cycles per unit time. Further increasing the hot bath temperature above a certain threshold does not increase the machine's power output. The cause is that, in this situation, the N-cycle has a greater probability than the H-cycle and R-cycle. Although the SSDB machine operates by randomly switching between  these three cycles, at the level of a single quantum jump trajectory, its heat engine efficiency and the refrigerator's coefficient of performance remain constant. 

\end{abstract}

\maketitle

\section{Introduction}
\label{section1} 

The Scovil and Schulz-DuBois (SSDB) quantum heat engine~\cite{Scovil1959} is the cornerstone of quantum thermodynamics~\cite{Alicki2018}. Since it was first proposed in 1959, the notion of quantum thermal machines has attracted considerable theoretical and experimental interest~\cite{Geusic1967,Alicki1976,Spohn1978,Spohn1978a, Kosloff1984,Geva1994,Feldmann2000,He2002,Scully2003,Quan2007,Boukobza2007,Segal2008,Allahverdyan2013,Su2016,Rossnagel2016,Agarwalla2017,Wang2017,Alicki2018,Peterson2019,Bouton2021,Xiao2023,Koch2023,Cangemi2024}.  

Although many quantum thermal machines have been proposed, owing to their simplicity and rich physics, SSDB machines continue to be investigated from various aspects ~\cite{Geva1994,Boukobza2007,Segal2018,Li2017,Singh2020,Kalaee2021,Menczel2021}. Figure~\ref{fig1}(a) schematically shows the machine. This system comprises a three-level atom and two heat baths with temperatures $T_1$ and $T_2$, where $T_1>T_2$. The atom is also weakly driven by a classically resonant single-mode field. Given that the foundation of the  machine is quantum physics, it essentially operates randomly and probabilistically. Earlier studies focused on the average behaviors or characteristics of the machine at the ensemble level~\cite{,Kosloff2013,Kosloff2014,Boukobza2007}. Nevertheless, following the rapid development of stochastic thermodynamics~\cite{Evans1993,Lebowitz1999,Esposito2009,Jarzynski2011,Seifert2011b,Campisi2011}, many efforts have been devoted to machine randomness~\cite{Li2017,Segal2018,Kalaee2021,Menczel2021}. For example, Menczel {\it et al.}~\cite{Menczel2021} and Kalaee {\it et al.}~\cite{Kalaee2021} applied full counting statistics to investigate violations of the thermodynamic uncertainty relation~\cite{Barato2015,Gingrich2016} in the quantum machine.

This paper arises from a simple question: Does the SSDB machine operate cyclically similar to classical thermal machines? Analogizing the quantum thermal machine to the classical machine is not new. In fact, this analogy was first introduced in the original paper of Scovil and Schulz-DuBois. They wrote that ``{\it As any heat engine, this system should be reversible so that it acts as a refrigerator. This is indeed the case. Suppose a quantum $h\nu_s$ is applied to the signal transition. It causes an ion} (the atom in this paper) {\it to go from state 1 to 2. The ion may further jump to state 3 if energy $h\nu_i$ is supplied by the cold reservoir. The cycle is finally completed when the ion returns from state 3 to state 1 while the energy $h\nu_p$ is communicated to the hot reservoir. In this process, energy is extracted from the idler transition, that is, from the cold reservoir, so that it is refrigerated.}". Similar statements are also present in recent papers~\cite{Li2017,Kalaee2021}. 
For example, Kalaee {\it et al.} extended the previous statement by considering randomness. They wrote that ``{\it Let $N$ denote the number of cycles done in the nominal direction {\it, i.e.}, $l\rightarrow x\rightarrow u\rightarrow l$} (the $l-$, $x-$, and $u$-levels corresponds to the quantum states $1$, $3$, and $2$ in this paper, respectively.) {\it minus the number of cycles in opposite direction. In each cycle, a photon is exchanged with each heat bath. }". 
This analogy is appealing because it provides an intuitive picture of the operation of the abstract quantum machine. However, it is also questionable. When we examine the previous statements, we do not see crucial terminologies of quantum physics, {\it e.g.}, wave function, superposition, and coherence. Hence, the analogy fails to reveal crucial physics in the quantum regime. These authors did not implement this analogy in practical calculations and analysis. 

In macroscopic thermodynamics, one cycle of a thermal machine is a thermodynamic process in which the system starts from an initial state and returns to the same initial state for the first time, {\it e.g.}, the famous Carnot cycle of a heat engine or refrigerator. An often-overlooked fact is that the cycles are executed on a single machine. Following the previous analogy, we might infer that the cycles in the SSDB machine would be perceived at the single-atom level if the cycle notion is valid. In the literature, this machine is typically described via the quantum master equation, which is related to the reduced density matrix of the atom~\cite{Gorini1976,Lindblad1976}. Hence, the theory is essentially an ensemble. In addition, it is well established that the quantum master equation can be unraveled into the quantum jump trajectory (QJT), a notion that is particularly relevant to the single-atom level ~\cite{Mollow1975,Srinivas1981,Scully1997,Breuer2002,Wiseman2010}. Therefore, in this paper, we attempt to explore the cycle notion in the trajectories and investigate the statistics and thermodynamic meanings of cycles in the operation of the SSDB machine. Although thermodynamic interpretations of QJTs have been recognized for many years~\cite{Breuer2004,DeRoeck2004,Horowitz2012,Hekking2013,Liu2014,Liu2016a}, to our knowledge, few studies have directly examined quantum thermal machines from the trajectory perspective~\cite{Liu2020,Menczel2020}.

The remainder of this paper is organized as follows. In Sec.~\ref{section2}, we unravel the quantum master equation of the SSDB machine into the QJT. In Sec.~\ref{section3}, we decompose the trajectory into a stochastic combination of three types of cycles. In Sec.~\ref{section4}, using a semi-Markov process method for open quantum systems, we investigate the means and fluctuation strengths of the cycle rates. The probabilities of the cycles are also exactly solved. In Sec.~\ref{section5}, we demonstrate that the heat engine's efficiency and the refrigerator's coefficient of performance defined in the QJTs are constant. Section~\ref{section6} concludes this paper.

\section{Unraveling the SSDB machine into quantum jump trajectories }
\label{section2}	
\begin{figure}
\includegraphics[width=1\columnwidth]{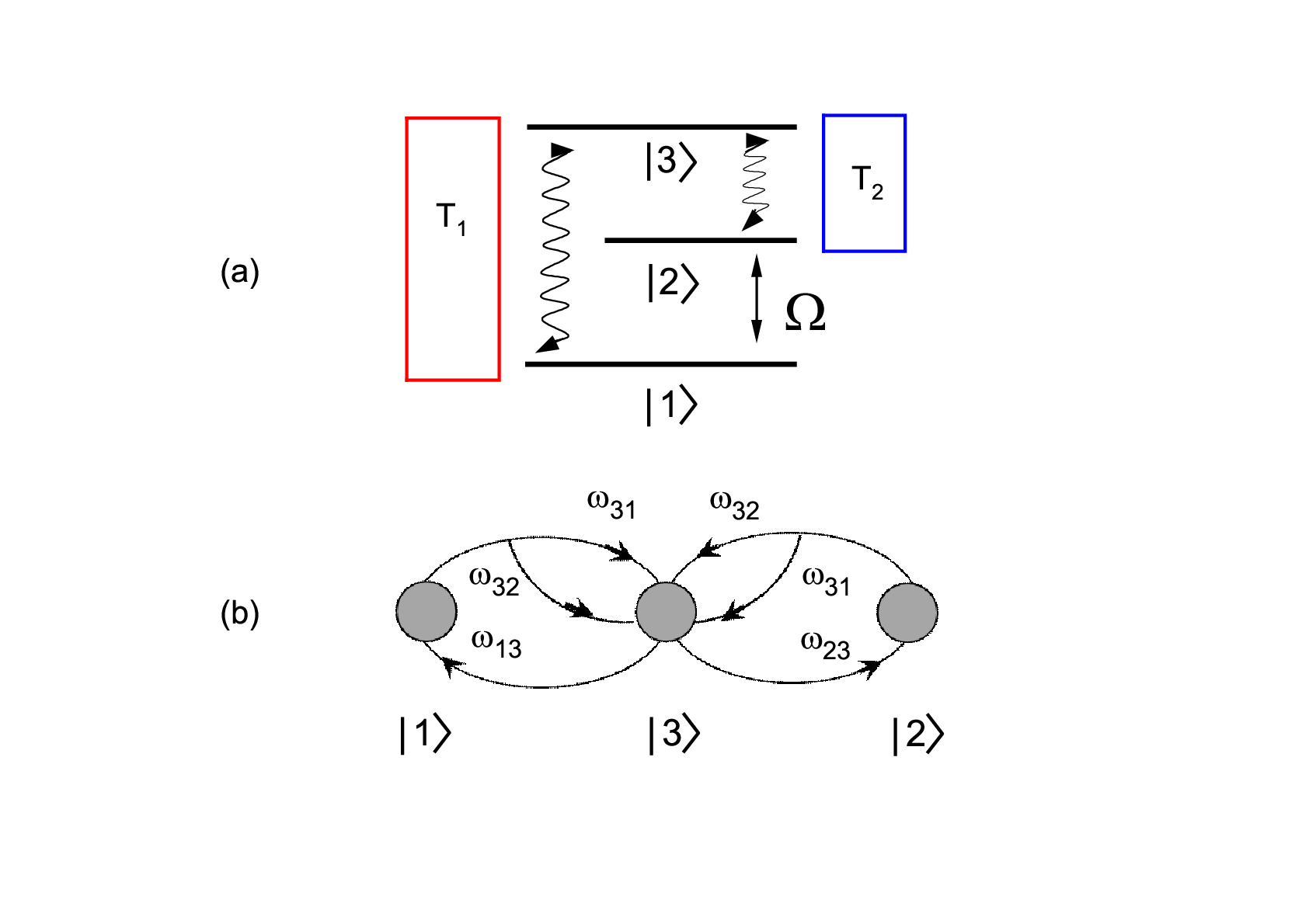}
\caption{(a) Schematic of the SSDB machine. The line with double arrows represents a classically resonant single-mode driving field. The coupling strength of the field with the three-level atom is $\Omega$. The squares on the left and right represent the heat baths, with temperatures $T_1$ and $T_2$, respectively. (b) Schematic of the QJTs of the SSDB machine. The gray dots indicate the collapsed quantum states. They are also the states from which the wave functions start. Curves without arrows represent deterministic pieces of the wave functions, whereas those with arrows represent collapses of the wave functions. Note that these curves merge in the trajectories connecting $|3\rangle$ and $|1\rangle$ or $|2\rangle$. The symbols $\omega_{3i}=E_3-E_i$ and $\omega_{i3}=E_i-E_3$, $i=1,2$, near the arrows, denote the quanta exchanged between the atom and the $T_i$-heat bath, where $E_i$ and $E_3$ are the energy levels of the atom. These quanta are also interpreted as the heat absorbed by the atom from the heat baths. A negative value of $\omega_{i3}$ indicates that the $T_i$-heat bath absorbs heat from the atom.}
\label{fig1}
\end{figure}
In the rotation framework, the quantum master equation for the three-level atom is  
\begin{eqnarray}
\label{QSSD}
	\partial_t\rho&=&-{\rm i}\left[ H,\rho(t)\right ] + {\cal D}_{13}[\rho]+{\cal D}_{23}[\rho].
\end{eqnarray}
In this equation, $\rho$ is the reduced density matrix of the atom, the Planck constant $\hbar$ is set to 1, and the interaction Hamiltonian between the atom and the field is   
\begin{eqnarray}
H=\Omega(\sigma_{12}+\sigma_{21}),
\end{eqnarray}
where $\Omega$ represents the Rabi frequency. We also assume that the field is resonant for simplicity in the calculations. Abandoning this assumption does not considerably change the main results in this paper. The action of the dissipated superoperator ${\cal D}_{i3}$, $i=1,2$, on the density matrix $\rho$ is as follows:    
\begin{eqnarray}
	\label{dissipatorsuperoperator}
{\cal D}_{i3}(\rho)&=&r_i^{-}\left[\sigma_{i3}\rho(t)\sigma_{3i} -\frac{1}{2}\{\sigma_{3i}\sigma_{i3}, \rho(t) \}   \right]\nonumber\\
&+&r_i^{+}\left[\sigma_{3i}\rho(t)\sigma_{i3} -\frac{1}{2}\{\sigma_{i3}\sigma_{3i}, \rho(t) \} \right], 
\end{eqnarray}
where $\sigma_{i3}=|i\rangle \langle 3|$ and  $\sigma_{3i}=|3\rangle \langle i|$ are the jump operators; $r_i^-=r_i(n_i+1)$ and  $r_i^+=r_in_i$ are the damping and exciting rates of the atom due to coupling with the $T_i$-heat bath, respectively; and $r_i$ and $n_i$ are the spontaneous emission rate and the Bose--Einstein distribution of the $T_i$ heat bath, respectively.   

Equation~(\ref{QSSD}) can be unraveled into the QJTs of single atoms~\cite{Mollow1975,Srinivas1981,Carmichael1993,Molmer93,Plenio1998,Breuer2002,Wiseman2010}. A trajectory is composed of deterministic pieces and random collapses of wave functions. Figure~\ref{fig1}(b) shows this scenario: The gray dots represent the quantum states $|i\rangle$, $i=1,2,3$, the curves without arrows denote that the wave functions of the  atom start from these states and continuously evolve, and the curves with arrows denote the random collapses of the wave functions into the states. From the perspective of a stochastic process, if only the collapse events of the wave functions are considered, which include the collapsed quantum states and collapse times, the QJTs can be considered realizations of a certain semi-Markov process~\cite{Liu2022,Liu2024a}. Under this idea, the evolution of the wave functions is replaced with the waiting time distributions of the stochastic process. For the SSDB machine, these distributions are formulated as   
\begin{eqnarray}
\label{WTDsformula1}	
p^{(j)}_{3|i}(\tau)&=&r_j^+ \left|\sigma_{3j}e^{-{\rm i}{\tilde H}\tau}|i\rangle\right|^2,\\
p_{i|3}(\tau)&=&r_i^-\left|\sigma_{i3}e^{-{\rm i}{\tilde H}\tau}|3\rangle\right|^2, 
\label{WTDsformula2}
\end{eqnarray}
$i,j=1,2$, where 
\begin{eqnarray}
\label{nonHermitianHamiltonian}
\tilde H=H-\frac{{\rm i}}{2}\sum_{i=1}^2r_i^-\sigma_{3i}\sigma_{i3}-\frac{{\rm i}}{2}\sum_{i=1}^2 r_i^+\sigma_{i3}\sigma_{3i} 
\end{eqnarray}
is the non-Hermitian Hamiltonian~\cite{Breuer2002}. 
In Eq.~(\ref{WTDsformula1}), $p^{(j)}_{3|i}$ is the probability density of the wave function starting from state $|i\rangle$, continuously evolving, and finally collapsing in state $|3\rangle$  through the jump operator $\sigma_{3j}$ at time $\tau$. In Eq.~(\ref{WTDsformula2}), $p_{i|3}$ has a similar probability density explanation except that the starting and collapsed states are $|3\rangle$ and $|i\rangle$, respectively. In the latter case, the jump operator is $\sigma_{i3}$. These formulas can be solved explicitly, and the expressions depend on the values of the physical parameters. Nevertheless, their Laplace transforms of time $\tau$ do not. We reserve Appendix A for these results. Here, we emphasize that $p_{i|j}$, $i,j=1,2$, are precisely zero. This result is determined by the special structure of Eq.~(\ref{QSSD}). Hence, the connections between the gray dots in Fig.~\ref{fig1}(b) are meaningful. This point is closely related to the operation of the SSDB machine. 
 
To support the above semi-Markov process perspective on the QJTs, we also conduct simulation of the trajectories. Because the computational algorithm is standard and detailed in the textbook~\cite{Breuer2002}, we do not explain it in this paper.  
 
\section{Three types of cycles}
\label{section3}
In this paper, we are interested in the energy exchanges between the three-level atom and the heat baths in the large time limit. According to a thermodynamic interpretation of the QJTs, a collapse of the wave function indicates a discrete amount of heat or a quantum absorbed by the atom from one of the heat baths. The sign and magnitude of the heat are given by the jump operators in Eq.~(\ref{QSSD})~\cite{Breuer2004,DeRoeck2004,Horowitz2012,Hekking2013,Liu2014,Liu2016a}.  We mark these quanta near the arrows in Fig.~\ref{fig1}(b). 
By simultaneously tracking the collapsed quantum states and the quanta, we may construct an energy ``path" and define thermodynamic quantities in an arbitrary trajectory. For example, let $\bf X$ be a QJT; then, the heat absorbed by the atom from the $T_1$- and $T_2$-heat baths is equal to
\begin{eqnarray}
\label{trajectoryheatT1}
Q_1[{\bf X}]&=&\omega_{31}\Delta m[{\bf X}], \\
\label{trajectoryheatT2}
Q_2[{\bf X}]&=&\omega_{32}\Delta n[{\bf X]}, 
\end{eqnarray}
respectively. Here, $\Delta m[\bf X]$ ($\Delta n[\bf X]$) denotes the net number of quanta $\omega_{31}$ ($\omega_{32}$) absorbed by the atom from the $T_1$- ($T_2$-) heat bath in the full trajectory, which is also equal to the number of quanta $\omega_{31}$ ($\omega_{32}$) minus the number of quanta  $\omega_{13}$ ($\omega_{23}$). Because of the first law of thermodynamics, the work output in the trajectory is  
\begin{eqnarray}
	\label{workquantumjumptrajectory}
	A[{\bf X}]=Q_1[{\bf X}]+Q_2[{\bf X}].  
\end{eqnarray}   

\begin{figure}
\includegraphics[width=1\columnwidth]{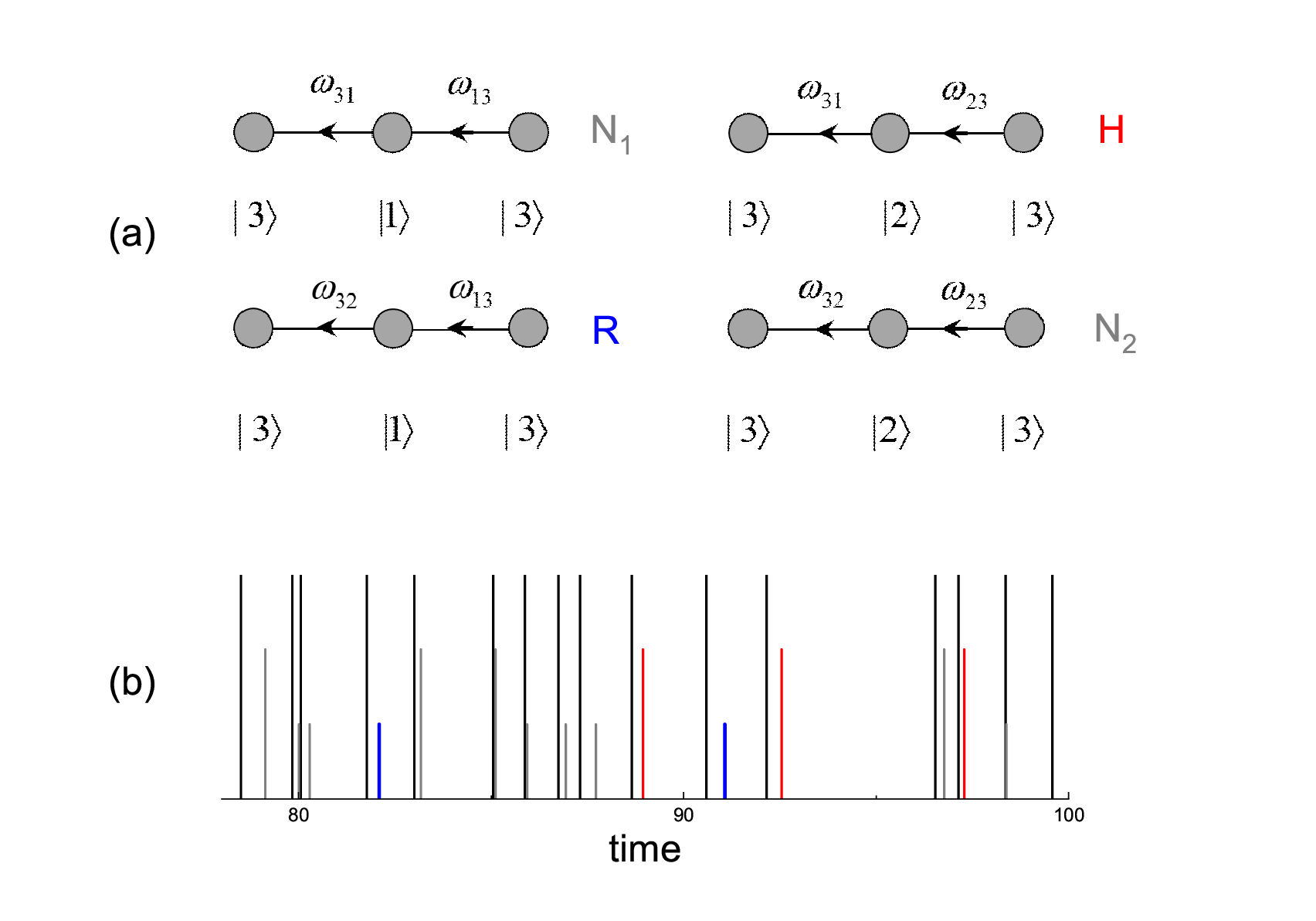}
\caption{(a) Diagrammatic representations of four types of cycles. They are uniquely designated by the specific collapsed quantum states and exchanged quanta. Note that we refer to the N$_1$-cycle and N$_2$-cycle as the same N-cycle for the remainder of this paper. (b) A simulated QJT of the SSDB machine. Long, medium, and short lines represent collapse times to states $|3\rangle$, $|2\rangle$, and $|1\rangle$, respectively. We observe that each short or medium line is always flanked by two long lines. Such a set of three lines composes a cycle. Based on the quanta in the trajectory, we further distinguish H-cycles, R-cycles, and N-cycles with red, blue, and gray lines, respectively. The simulation parameters are $\omega_{31}=3$, $\omega_{32}=1$, $r_{1}=r_{2}=1$, $n_1=1$, $n_2=0.5$, and $\Omega=0.5$.   }
\label{fig2}
\end{figure}

After carefully inspecting Fig.~\ref{fig1}(b), we notice that a QJT is a stochastic combination of cycles~\footnote{In this paper, we neglect noncycle parts of the trajectories, if any, at the beginning and ending of the trajectories, which is reasonable considering the large time limit. }. Here, a cycle refers to a short trajectory that starts from state $|3\rangle$ and ends in the same state for the first time and we denote it $\bf C$. Considering the accompanying quanta in these cycles, we further classify them into four types: N$_1$, R, H, and N$_2$. Figure~\ref{fig2}(a) presents their diagrammatic representations. Note that we depict these quantum states from right to left. The energy scenarios of these cycles are straightforward. We observe that in any type of cycle, the identity   
\begin{eqnarray}
	\label{identitydmdn}
	\Delta m[{\bf C}]=-\Delta n[{\bf C}]
\end{eqnarray}
holds, where $\Delta m[\bf C]$ is simply equal to $\pm 1$ or $0$. Equation~(\ref{identitydmdn}) reminds us of a statement made by Mitchison~\cite{Mitchison2019}. He argued that there is a perfect cooperative transfer of precisely one quantum each from the cold heat bath into the hot one. Our identity~(\ref{identitydmdn}) not only extends this statement to the context of cycles but also covers a reverse procedure and an overlooked case involving zero quantum transfer. If we apply Eqs.~(\ref{workquantumjumptrajectory}) and~(\ref{identitydmdn}) to the cycles, the work output is   
\begin{eqnarray}
A[{\bf C}]%\omega_{31}\Delta m[{\bf C}]+\omega_{32}\Delta n[{\bf C}]
=\omega_{21}\Delta m[{\bf C}]. 
\end{eqnarray} 
The physical relevance of these four types of cycles is now apparent: The R-cycle indicates that the SSDB machine operates in the cycle as a refrigerator by work input ($\Delta m[{\bf C}]=-1$) and transferring a quantum $\omega_{32}$ from the cold heat bath to the hot one; in the H-cycle, the machine operates as a heat engine by absorbing a quantum $\omega_{31}$ from the hot heat bath and a work output ($\Delta m[{\bf C}]=1$); and in the N$_1$- and N$_2$-cycles, the machine is idle or neutral ($\Delta m[{\bf C}]=0$). Owing to their identical vanishing operations, we refer to the N$_1$-cycle and the N$_2$-cycles as N-cycles for the remainder of this paper and do not discuss them separately. Figure~\ref{fig2}(b) clearly illustrates these results in a simulated QJT.

Let us return to the arbitrary QJT $\bf X$. On the basis of the above discussion, we conclude that the SSDB machine operates randomly by switching among three distinct types of cycles. Accordingly, the thermodynamic quantities possess alternative expressions that are based on the cycles: 
\begin{eqnarray}
\label{cyclicheatT1}
Q_1[{\bf X}]&=&\omega_{31}\sum_{i}\Delta m[{\bf C}_i], \\
\label{cyclicheatT2}
Q_2[{\bf X}]&=&\omega_{32}\sum_{i}\Delta n[{\bf C}_i],\\
\label{cyclicwork}
	A[{\bf X}]&=&\omega_{21}\sum_i\Delta m[{\bf C}_i].
\end{eqnarray}
Here, the subscript $i$ denotes the $i$-th cycle in the trajectory.   
Although the SSDB machine is stochastic, Eq.~(\ref{workquantumjumptrajectory}) implies that we can still define it as a heat engine or refrigerator if the work output is positive or negative in the large time limit. In this context, studying rates instead of time-extensive quantities is more convenient~\cite{Touchette2008}. To this end, we define power ${\cal P}=A[{\bf X}]/t$ and the net number of quanta $\omega_{31}$ absorbed by the atom from the $T_1$-heat bath per unit time,
\begin{eqnarray}
\label{countingnetrateT1bath}
{\cal M}=\frac{\Delta m[{\bf X}]}{t}=\frac{\sum_i \Delta m[{\bf C}_i] }{t}. 
\end{eqnarray}
These two quantities are random variables that are proportional to each other. In addition, because $\Delta m[\bf C]$ is equal to $-1$ and $+1$ for the R-cycle and H-cycle, respectively, $\cal M$ has a  further expression:   
\begin{eqnarray}
\label{RandHrates} 
{\cal M}&=&\frac{\#[\text{H-cycles}]}{t}-\frac{\#[\text{R-cycles}]}{t}\nonumber\\ 
&=&{\cal C}_H-{\cal C}_R, 
\end{eqnarray}
where $\#[\cdots]$ represents ``the number of $\cdots$ ". In Eq.~(\ref{RandHrates}), ${\cal C}_R$ and ${\cal C}_H$ are the rates of the SSDB machine operating as a refrigerator and heat engine, respectively. We simply call them the R-cycle and H-cycle rates. The new expression of $\cal M$ indicates that in the large time limit a competition between the R-cycle and H-cycle rates determines the operation of the machine. In the following section, we focus on the statistics of these two rates.    

\section{Statistics of cycles}
\label{section4}
\subsection{Mean and fluctuation strength of cycle rates}
A careful inspection of Fig.~\ref{fig1}(b) indicates that a QJT in which the atom starts from state  $|1\rangle$ and evolves continuously until it collapses into state $|3\rangle$ with the absorption of quantum $\omega_{32}$ uniquely specifies the R-cycle. A similar observation applies to the H-cycle. Therefore, the statistics of the two cycles correspond to the counting statistics of these two specific trajectories. Taking the R-cycle as an example, its statistical properties are obtained by solving the scaled cumulant generating function $\varphi(\lambda)$, which is the largest real root of the algebraic equation~\cite{Liu2022,Liu2024a}, 
\begin{eqnarray}
	\label{matrixeqforcountingstat}
	\hat p_{1|3}(\nu)\left(\hat p_{3|1}^{(1)}(\nu) + \hat p_{3|1}^{(2)}(\nu)e^\lambda  \right)+  
	\hat p_{2|3}(\nu)\left(\hat p_{2|3}^{(1)}(\nu)+ \hat p_{2|3}^{(2)}(\nu)\right)=1,
\end{eqnarray}
where $\nu$ is complex frequency of the Laplace transforms of the waiting time distributions, and $\lambda$ is the parameter of the cumulant generating function. An explanation of this equation is provided in Appendix B. Substituting the concrete waiting time distributions, we see that Eq.~(\ref{matrixeqforcountingstat}) reduces to a quartic equation. Its radical solutions are complicated. Because the scaled cumulant generating function provides much statistical information, here, we focus on the mean $\overline{{\cal C}}_R$ and fluctuation strength $F_R$ of the R-cycle rate. Aided by Eqs.~(\ref{mean}) and~(\ref{variance}), we solve 
\begin{eqnarray}
\label{meanR}
\overline{{\cal C}}_R&=&
	\frac{2\Omega^2}{D}r_1^-r_2^+,\\
\label{varianceR}
F_R&=&\lim_{t\rightarrow\infty}t\text{Var}[{\cal C}_R]=\overline{{\cal C}}_R-\frac{4E\Omega^4}{D^3}(r_1^-r_2^+)^2, 
\end{eqnarray}
where $D$ and $E$ are  
\begin{eqnarray}
D&=&2\Omega^2\left[ r_1^++r_2^++2(r_1^-+r_2^-)\right]+\frac{1}{2}(r_1^++r_2^+)\left(r_1^+r_2^++r_1^+r_2^-+r_1^-r_2^+\right),\\
E&=&8\Omega^2+(r_1^+)^2+(r_2^+)^2+4r_1^+r_2^++r_1^+r_1^-+r_2^+r_2^-+3(r_1^+r_2^-+r_1^-r_2^+),
\end{eqnarray}
respectively. Analogously, we solve the mean and fluctuation strength of the H-cycle rate. However, owing to the symmetry in the waiting time distributions, these two formulas can be derived by exchanging indices $1$ and $2$ in Eqs.~(\ref{meanR}) and (\ref{varianceR}). Note that $D$ and $E$ are invariant under this operation. Figure~\ref{fig3}(a) shows the statistical quantities of the two rates under a set of parameters, in which the distribution $n_1$ is varied and $n_2$ is fixed at a certain value. The features of the mean rate curves are similar to those of the fluctuation strength curves in each type of cycle. In addition, the intersection points of these curves occur at $n_1=n_2$, which are apparent according to Eqs.~(\ref{meanR}) and~(\ref{varianceR}). To verify these analytical formulas, we conduct a QJT simulation to obtain the same quantities, as represented by the symbols in Fig.~\ref{fig3}(a). We clearly see that the data excellently agree with the formulas.

Let us examine Fig.~\ref{fig3}(a). First, the intersection point of the mean rate curves denotes a transition of the operation of the SSDB machine from a refrigerator to a heat engine. The reason is explained in Eq.~(\ref{RandHrates}): By taking the time average of the equation, we obtain  
\begin{eqnarray}
\label{avgcountingnetquantaT1bath}
\overline{\cal M}=\frac{2\Omega^2}{D}(r_1^+r_2^--r_1^-r_2^+)\propto (n_1-n_2), 
\end{eqnarray}
and the power output $\cal P$ is proportional to $\cal M$. Here, a ``bar" placed over a symbol denotes a mean in time. The transition condition $n_1=n_2$ is consistent with that derived at the ensemble level~\cite{Boukobza2007}. Second, when the SSDB operates as a heat engine, a further increase in the population $n_1$ above a threshold value decreases the output power. We mark the threshold value with an arrow in the figure. A similar observation is presented in the fully quantum SSDB model~\cite{Li2017}. Because the energy levels of the atom are considered given, changes in $n_1$ are equivalent to changes in the temperature $T_1$ of the hot bath. We expect that, with the temperature of the cold bath held constant, the higher the temperature of the hot bath is, the greater the power output.

\subsection{Probabilities of cycles}
To understand the counterintuitive second result, we revisit the R-cycle and H-cycle rates. Taking the R-cycle as an example, in addition to the definition given in Eq.~(\ref{RandHrates}), its rate has an alternative expression:     
\begin{eqnarray}
\label{probexpressionofrateofRcycle}
{\cal C}_R&=&\frac{\#[\text{cycles}]}{t}{\times } 	\frac{\#[\text{R-cycles}]}{\#[{\rm cycles}]}. 
\end{eqnarray}
The first term on the right-hand side of Eq.~(\ref{probexpressionofrateofRcycle}) represents the number of cycles per unit time. We call this property the cycle rate and denote it as $\cal C$. The second term in Eq.~(\ref{probexpressionofrateofRcycle}) is the ratio of the number of R-cycles to the total number of cycles in a QJT. Hence, it approximates the probability $P_R$ of a cycle being the R-type. In the  large time limit, the approximation becomes exact. Therefore, Eq.~(\ref{probexpressionofrateofRcycle}) reduces to $\overline{\cal C}_R=\overline{\cal C} P_R$. In terms of the Laplace transforms of the waiting time distributions, the  probability is equal to  
\begin{eqnarray}
\label{probRcycle}
P_R=\hat p^{(2)}_{31}(0)\hat p_{13}(0). 
\end{eqnarray}
For the H-cycle rate, we have an analogous equation, $\overline{\cal C}_H=\overline{\cal C} P_H$, where the probability of a cycle being H-type is 
\begin{eqnarray}
\label{probHcycle}
P_H=\hat p^{(1)}_{32}(0)\hat p_{23}(0). 
\end{eqnarray}  
Finally, there is an N-cycle rate, ${\cal C}_N$. Its mean is $\overline{\cal C}_N=\overline{\cal C}P_N$, where $P_N=1-P_H-P_R$ is the probability of a cycle being neutral. 

%Note that if the N-cycle rate were absent, we would obtain an incorrect identity:   
%\begin{eqnarray}
% \overline{\cal C}_R+\overline{\cal C}_H=\overline{\cal C}.
%\end{eqnarray} 
 
\begin{figure}
\includegraphics[width=1\columnwidth]{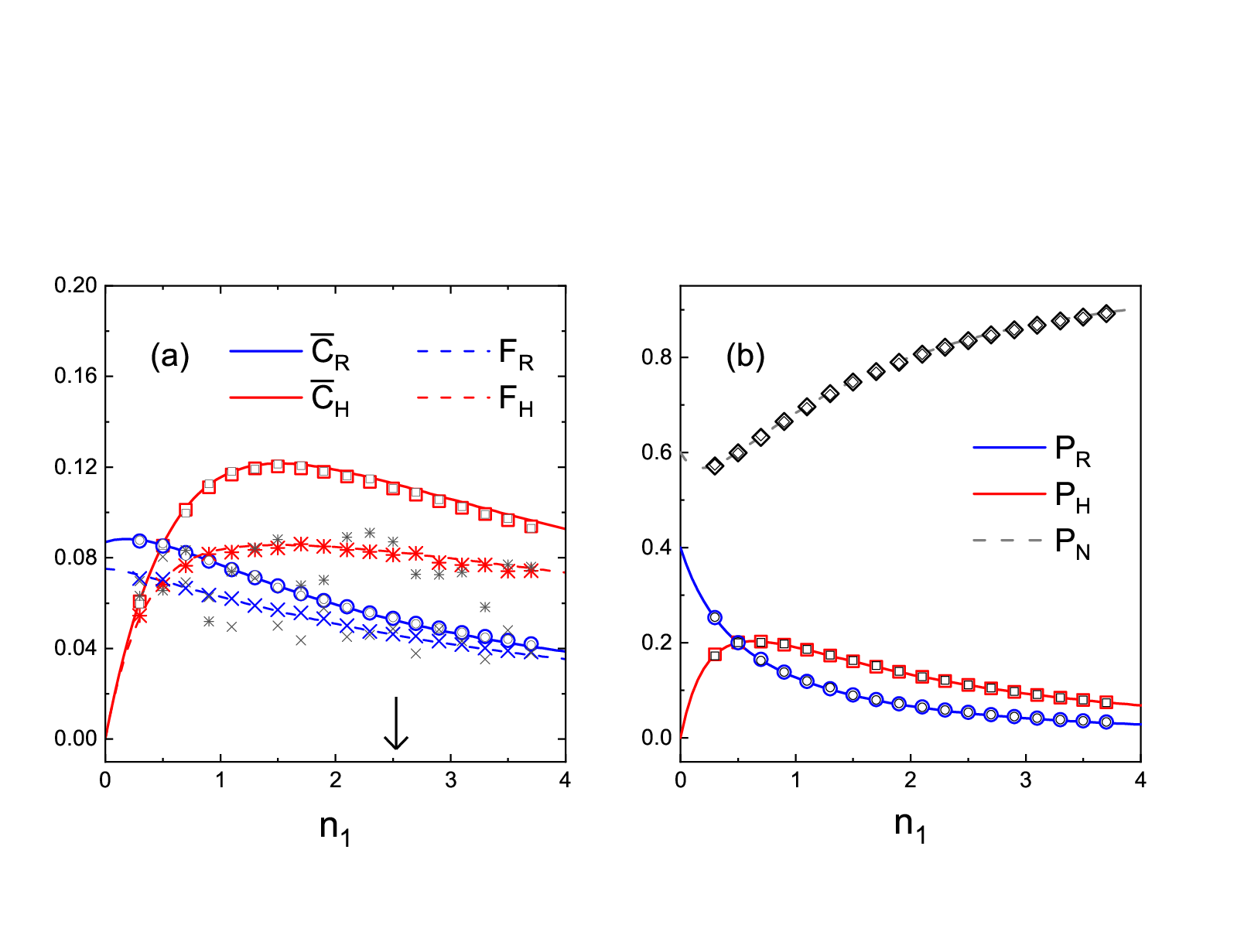}
\caption{(a) Mean rates and fluctuation strengths of the R-cycle and H-cycle rates. The arrow denotes the threshold value at which the power output of the SSDB machine starts to decrease if $n_1$ further increases. Note that the mean H-cycle rate has already exceeded its maximum and starts to decrease before reaching the threshold value. (b) Probabilities that cycles are H, R, and N cycles. In both panels, the $x$-axis is the population $n_1$. The fixed parameters are $\omega_{31}=3$, $\omega_{32}=1$, $r_{1}=r_{2}=1$, $n_2=0.5$, and $\Omega=0.5$. Because we set $T_1>T_2$, the minimum value of $n_1$ is given by  $1/((1 + 1/n_2)^{w31/w32}- 1)=0.0384$. The solid and dashed curves are calculated by the analytical formulas. %, Eqs.~(\ref{meanR}), (\ref{varianceR}), (\ref{probRcycle}), and (\ref{probHcycle}). 
The symbols, which include the open squares, open circles, stars, and crosses, are the data calculated via simulating QJTs. The simulation duration is 1000. The number of trajectories for symbols with small sizes is 100, while for symbols with large sizes it is 10,000. We observe that the convergence of the mean rates and probabilities of the cycles is significantly faster than that of the fluctuation strengths. }
\label{fig3}
\end{figure}

We derive the mean cycle rate via Eq.~(\ref{meanR}) and the Laplace transforms of Eq.~(\ref{LaplacetransformsWTDs}):  
\begin{eqnarray}
\label{meancyclerate}
\overline{\cal C}=\frac{1}{2D}(4\Omega^2+r_1^+r_2^+)(r_1^++r_2^+)(r_1^-+r_2^-). 
\end{eqnarray}  
Note that this value is also equal to half of the mean activity~\cite{Garrahan2017}, which is explained in Appendix B. The data show that Eq.~(\ref{meancyclerate}) is a monotonically increasing function of $n_1$.  Consequently, in Fig.~\ref{fig3}(a), the observed nonmonotonic feature of the $\overline{\cal C}_H$ curve is attributed to the probability $P_H$. To illustrate this aspect, we depict $P_R$, $P_H$, and $P_N$ under the same set of parameters in Fig.~\ref{fig3}(b). We note that the probability $P_N$ strongly suppresses the other two probabilities when the population $n_1$ is greater than $n_2$. Thus, we obtain the following conclusion. With larger $n_1$ 
values or at higher temperatures $T_1$, the SSDB machine becomes more active or operates at a higher cyclic frequency. However, the rapid increase in the number of N cycles leads to a corresponding decrease in the number of the other two cycle types. The simulation data also verify this analysis; see the symbols in Fig.~\ref{fig3}(b).

We close this section by mentioning that the statistics of $\cal M$ can be directly derived through its original definition, the first equation in Eq.~(\ref{countingnetrateT1bath}), without resorting to Eq.~(\ref{RandHrates}). Kalaee et al. derived its mean $\overline{\cal M}$ and fluctuation strength via the full counting statistics~\cite{Kalaee2021}. Appendix B provides an alternative derivation via the semi-Markov process method. Although $\overline{\cal M}$ must be the same as that in Eq.~(\ref{avgcountingnetquantaT1bath}), the equation now has a cyclic explanation. Another informative formula is the fluctuation strength 
\begin{eqnarray}
	\label{Mcovariance}
	F_{\cal M}&=&\overline {\cal C}_H+\overline {\cal C}_R-\frac{4E\Omega^4}{D^3}(r_1^+r_2^--r_1^-r_2^+)^2\\
	&=&F_H+F_R+\frac{8E\Omega^4}{D^3}r_1^+ r_1^- r_2^+ r_2^-. 
	\label{McovariancebyRHcyclecovariances} 
\end{eqnarray}
Equation~(\ref{Mcovariance}) is derived by  the original definition. We emphasize that  this equation cannot be obtained from the definition in Eq.~(\ref{RandHrates}).  Equation~(\ref{McovariancebyRHcyclecovariances}) is a simple consequence of substitutions of $F_R$ and $F_H$. Over a large but finite duration, Eq.~(\ref{McovariancebyRHcyclecovariances}) implies that the covariance of ${\cal C}_H$ and ${\cal C}_R$ is nonzero and negative. Hence, the H-cycle and R-cycle rates are negatively correlated. The reason is understandable. A cycle is either a heat engine, refrigerator, or neutral. If the likelihood of a cycle being heat engine increased, the other likelihood including being refrigerator would decrease.  
 
\section{Efficiency and coefficient of performance}
\label{section5}
We demonstrate that the heat engine's efficiency and the refrigerator's coefficient of performance of the SSDB machine are constant rather than fluctuating. In the large time limit, given a QJT $\bf X$, they are   
\begin{eqnarray}
\label{efficiency}
\eta_e&=&\frac{A[\bf X]}{Q_1[\bf X]}=\frac{\omega_{21}}{\omega_{31}},
	\label{SSDB's efficiency}\\
\label{COP}
c_p&=&-\frac{Q_2[\bf X]}{A[{\bf X}] }=\frac{\omega_{32}}{\omega_{21}},
\end{eqnarray}
respectively. Here, we substituted Eqs.~(\ref{cyclicheatT1})-(\ref{cyclicwork}) into Eqs.~(\ref{efficiency}) and~(\ref{COP}), respectively. Intriguingly, Eq.~(\ref{SSDB's efficiency}) is precisely the SSDB's maser efficiency formula~\cite{Scovil1959}.

\section{Conclusions} 
\label{section6}
In this paper, we explore cycles in the QJTs of the SSDB machine. Although the quantum thermal machine is often thought of as operating cyclically, the explicit cyclic structure and precise thermodynamic meaning of the cycles are ambiguous. In the trajectories, we identify cyclic structures that can be classified into three types: H-, R-, and N-cycles. Accordingly, in each cycle, the machine operates as a heat engine, refrigerator, or neutral. In a trajectory with a long duration, the SSDB machine operates by randomly switching between these three cycle types. We emphasize that the H-cycle and R-cycle are not opposite processes, since each cycle has two states, but one of the two states is not the same. Therefore, statements used in classical thermal machines, such as completing one cycle in the three states in sequence and concepts of clockwise (forward) and counterclockwise (backward) cycles, are inappropriate in the quantum case. In addition to the notion and picture, we also study the statistical properties of these three cycles via the semi-Markov process method. We find that the power output of the SSDB machine actually decreases if the hot bath temperature exceeds a certain threshold value. From a cyclic perspective, this decrease is attributed to the probability of the machine being neutral surpassing the probabilities of the other two cycle types. Note that the mean cycle rate actually increases under this situation. We finally demonstrate that the heat engine's efficiency and the refrigerator's coefficient of performance are constant in a single QJT. This constancy results from the unique structure of the SSDB machine. 

{\it Note added}. After completing this work, Abhaya S. Hegde informed us that his group has obtained a very similar conclusion that the dynamics of the three-level atom (maser) can be resolved into cycles that are classified as engine-like, cooling-like, or idle (arXiv:2408.00694v2). Significantly, their paper clarifies that a quantum thermal machine, operating randomly among the three types of cycles, requires its quantum master equation to adhere to specific mathematical restrictions.
\\
\\{\noindent\it Acknowledgments}
We are grateful for the discussions on quantum heat engines with Shihao Xia and Prof. Shanhe Su during this work. We also express our gratitude to Prof. Jianhui Wang for his inspiring discussion on stochastic efficiency. This work was supported by the National Natural Science Foundation of China under Grant No. 12075016 and No. 11575016.\\

\appendix
\section{Laplace transforms of the waiting time distributions}
According to Eqs.~(\ref{WTDsformula1}) and ~(\ref{WTDsformula2}) and the non-Hermitian Hamiltonian~(\ref{nonHermitianHamiltonian}), we calculate the Laplace transforms of the waiting time distributions with respect to time:   
\begin{eqnarray}
\label{LaplacetransformsWTDs}
\hat p_{3|1}^{(1)}(\nu)&=&r_1^+\frac{\nu^2-\nu(h_++h_--r_2^+)+2h_+h_-+(r_2^+)^2/2 }{(\nu-2h_+)(\nu-2h_-)[\nu-(h_++h_-)] },\nonumber\\
\hat p_{3|1}^{(2)}(\nu)&=&r_2^+\frac{2\Omega^2}{(\nu-2h_+)(\nu-2h_-)[\nu-(h_++h_-)]},\\
\hat p_{1|3}(\nu)&=&r_1^-\frac{1 }{\nu+(r_1^-+r_2^-)}, \nonumber
\end{eqnarray}
where $\nu$ is complex frequency of the Laplace transform. Other expressions that include $\hat p_{3|2}^{(1)}$, $\hat p_{1|3}^{(2)}$, and $\hat p_{2|3}$ are obtained by exchanging indices $1$ and $2$ in Eq.~(\ref{LaplacetransformsWTDs}), {\it e.g.},
\begin{eqnarray}
\hat p_{3|2}^{(2)}(\nu)&=&r_2^+\frac{\nu^2-\nu(h_++h_--r_1^+)+2h_+h_-+(r_1^+)^2/2 }{(\nu-2h_+)(\nu-2h_-)[\nu-(h_++h_-)] }.\nonumber
\end{eqnarray}
Here, $h_+$ and $h_-$ are two roots of a quadratic equation: $h_++h_-=-(r_1^++r_2^+)/2$ and $h_+h_-=\Omega^2+r_1^+r_2^+/4$. 
Symmetry arises from the resonant condition in which the detuning parameter is zero.  
When the discriminant $(r_1^+-r_2^+)^2-16\Omega^2$ is positive or negative, the two roots can be both real numbers or complex conjugates to each other. Correspondingly, in the time domain, $p^{(j)}_{3|i}$ and $i,j=1,2$ are  multiexponential decay and oscillatory decay, respectively.  Equation~(\ref{LaplacetransformsWTDs}) has a useful property: At $\nu=0$, these transforms are just the transition probabilities of the Markov chain in the semi-Markov process~\cite{Liu2022}.      
 
\section{Semi-Markov process method} 
In the main text, we noted that the statistics of the R-cycle are equivalent to the counting statistics of the specific QJT. Let the random variable ${\cal C}_R$ have the distribution $p_R(c)$. Its statistics can be obtained via finding the cumulant generating function, the logarithm of the Laplace transform of the distribution. In the large time limit, the scaled cumulant generating function      
\begin{eqnarray}
\varphi(\lambda)=\lim_{t\rightarrow\infty}\frac{1}{t}\ln\langle e^{\lambda tc}\rangle 
\end{eqnarray}
is more useful~\cite{Touchette2008}, where $\langle\cdots\rangle$ denotes an average over the distribution. The semi-Markov process method can solve the functions in open quantum systems~\cite{Liu2022,Liu2024a}. The core of the method is that the scaled cumulant generating function of ${\cal C}_R$ is equal to the largest real root of the equation of poles:   
\begin{eqnarray}
	\label{equationofpoles}
	\det[{\mathbb I}-{\mathbb W}(\nu,\lambda)]=0.
\end{eqnarray}
where $\mathbb{I}$ is a $3\times 3$ unit matrix and 
\begin{eqnarray}
\mathbb{W}(\nu,\lambda)=
\begin{bmatrix}
		0&0&\hat p_{1|3}(\nu) \\
		0 &0&\hat p_{2|3}(\nu) \\
		\hat p^{(1)}_{3|1}(\nu)+\hat p^{(2)}_{3|1}(\nu)e^\lambda &\hat p^{(1)}_{3|2}(\nu)+\hat p^{(2)}_{3|2}(\nu)  & 0. \\
\end{bmatrix}
\end{eqnarray}
Here, the structure of the matrix $\mathbb W$ already considers the characteristics of the waiting time distributions in the SSDB machine. The semi-Markov process method has been described previously~\cite{Liu2022,Liu2024a}. Here, we summarize a rule for writing the matrix $\mathbb W$: When a specific QJT is counted, in which the atom starts in quantum state $|i\rangle$, $i=1,2$, continuously evolves, and first collapses into state $|3\rangle$ by a certain jumping operator $\sigma_{3j}$, $j=1,2$, the Laplace transform of $\hat p^{(j)}_{3|i}$ in the matrix is multiplied with $\exp(\lambda)$; when another specific trajectory is counted, in which the atom starts $|3\rangle$, continuously evolves, and first collapses into state $|i\rangle$, $i=1,2$, the Laplace transform $\hat p_{i|3}$ is multiplied by $\exp(\lambda)$.

In general, the equation of poles~(\ref{equationofpoles}) reduces to a higher-order algebraic equation. Hence, finding its roots analytically is usually difficult. However, the first two moments can be derived analytically. First, we transform Eq.~(\ref{equationofpoles}) into a polynomial in $\nu$ by removing all denominators and let the left-hand side be $P(\nu,z)$ with $z=\exp(\lambda)$. Following the definition of the CGF, we have 
\begin{eqnarray}
\left. P(\nu,z)\right|_{\nu=0,z=1}=0. 
\end{eqnarray} 
Second, by taking derivatives up to second order, we derive  
\begin{eqnarray}
	\label{mean}
	\frac{d\nu}{d\lambda}(0)&=&-\left.\frac{ \partial_z P}{\partial_\nu P}\right|_{\nu=0,z=1}, \\
	\frac{d^2\nu}{d\lambda^2}(0)&=&-\left.\frac{1}{\partial_\nu P} \left[  \left(\frac{d\nu}{d\lambda}\right)^2\partial_\nu^2 P+2 \left(\frac{d\nu}{d\lambda}\right) \partial_{\nu z}^2 P + \partial_z^2 P + \partial_z P  \right]\right|_{\nu=0,z=1}. 
	\label{variance}
\end{eqnarray}
Finally, considering $\nu$ as the scaled cumulant generating function $\varphi(\lambda)$, Eqs.~(\ref{mean}) and~(\ref{variance}) are the mean and fluctuation strength of the R-cycle rate, respectively. The procedure for calculating the statistics of the R-cycle rate also applies to the H-cycle rate. The only modification is to shift the position of $e^\lambda$ from after $\hat p^{(2)}_{3|1}$ to after $\hat p^{(1)}_{3|2}$. 

We still need to calculate the counting statistics for ${\cal C}$, the cycle rate, and $\cal M$, the net number of quanta $\omega_{31}$ absorbed by the atom from the $T_1$-heat bath per unit time.  Similarly, modifications are made to matrix $\mathbb W$. For $\cal C$, Fig.~\ref{fig2}(b) indicates that a cycle is completed when specific QJTs that start from state $|3\rangle$ and end in state $|1\rangle$ or $|2\rangle$ for the first time are counted.  Accordingly, the matrix is modified as follows:     
\begin{eqnarray}
\label{Wmatrixcyclerate}
\mathbb{W}(\nu,\lambda)=	
\begin{bmatrix}
		0&0&\hat p_{1|3}(\nu) e^\lambda  \\
		0 &0&\hat p_{2|3}(\nu)e^\lambda \\
		\hat p^{(1)}_{3|1}(\nu) +\hat p^{(2)}_{3|1}(\nu)  
		&\hat p^{(1)}_{3|2}(\nu)  +\hat p^{(2)}_{3|2}(\nu)  & 0 \\
\end{bmatrix}.
\end{eqnarray}
A comment is in order. Reference~\cite{Garrahan2017} proposed the notion of activity. In the context of the trajectories, activity refers to the number of wave functions that collapse per unit time. Its statistics are derived from another $\mathbb W$, in which all Laplace transforms of the waiting time distributions are multiplied by $\exp(\lambda)$. The equation of poles for the newly mentioned matrix is identical to that of Eq.~(\ref{Wmatrixcyclerate}) if the $\lambda$ in the latter is replaced by $2\lambda$. Thus, we mathematically demonstrate that the mean activity is twice the mean cycle rate. Since two collapses occur in any cycle, this conclusion is easily understood. For the random variable ${\cal M}$, the matrix is  
\begin{eqnarray}
	\label{matrixforM}
\mathbb{W}(\nu,\lambda)=	
\begin{bmatrix}
		0&0&\hat p_{1|3}(\nu)e^{-\lambda}  \\
		0 &0&\hat p_{2|3}(\nu) \\
		\hat p^{(1)}_{3|1}(\nu) e^{\lambda}+\hat p^{(2)}_{3|1}(v) 
		&\hat p^{(1)}_{3|2}(\nu)e^{\lambda}  +\hat p^{(2)}_{3|2}(\nu)& 0 \\
\end{bmatrix}.
\end{eqnarray}
Note that $\exp(-\lambda)$ after $\hat p_{1|3}(\nu)$ in Eq.~(\ref{matrixforM}) indicates that the $T_1$-heat bath absorbs a quantum $\omega_{31}$ from the atom rather than the reverse process. 

We briefly explain why we chose the semi-Markov process method instead of the more popular full counting statistics~\cite{Mollow1975,Lebowitz1999,Bagrets2003,Esposito2009,Garrahan2010,Landi2024}. To solve the scaled cumulant generating functions, the latter method similarly involves finding the largest real eigenvalue of the tilted generator of the quantum master equation. We have demonstrated that these two methods are equivalent when calculating the statistics of certain special random variables, such as those previously mentioned $\cal C$ and $\cal M$. However, the full counting statistics is unsuited for the H-cycle and R-cycle rates that are of interest in this paper. For more explanations, we refer to our previous paper~\cite{Liu2022}.    
 
%\bibliography{RFsubmission20240730}

\end{document}